\documentclass[twocolumn,aps,prl,showpacs]{revtex4}
\usepackage{graphicx}
\usepackage{dcolumn}
\begin{document}

\title{Short-range type critical behavior in spite of long-range 
interactions:
\linebreak the phase transition of a Coulomb system on a lattice}

\author{A.~M\"obius\footnote{e-mail: a.moebius@ifw-dresden.de} 
and U.K.~R\"o{\ss}ler}

\affiliation{Leibniz Institute for Solid State and Materials Research 
       Dresden, PF 27 01 16, D-01171 Dresden, Germany}
       
\date{\today}

\begin{abstract}     
One- to three-dimensional hypercubic lattices half-filled with localized 
particles interacting via the long-range Coulomb potential are 
investigated numerically. The temperature dependences of specific heat, 
mean staggered occupation, and of a generalized susceptibility indicate 
order-disorder phase transitions in two- and three-dimensional systems. 
The critical properties, clarified by finite-size scaling analysis, are 
consistent with those of the Ising model with short-range interaction.
\end{abstract} 


\maketitle

The problem under which conditions Coulomb glasses exhibit genuine 
phase transitions has been under controversial debate for two decades
\cite{DLR82,V93,GY93,VS94,Detal00}. The significance of the kind of 
static disorder is unclear yet \cite{GY93,VS94}. In this context, the 
relation to the Ising model with short-range interaction is of great 
interest. It would be very useful to know how replacing its 
nearest-neighbor coupling by an ``antiferromagnetic'' long-range 
Coulomb interaction modifies the critical behavior of a system without 
static disorder. Due to the interplay of long-range interaction and 
frustration, one might alternatively expect this model to belong to
different universality classes: In case of a sufficiently slowly 
decaying ferromagnetic power-law interaction, including the 
$1/r$-interaction, mean field behavior is obtained \cite{LB02}. But the 
condensation in a three-dimensional model of an electrolyte exhibits 
Ising-like critical behavior, possibly due to efficient screening 
\cite{Letal02}.

To decide the question, we numerically study lattices half-filled with 
localized particles interacting via the long-range Coulomb potential,
\begin{equation}
H = \frac{1}{2} \sum_{i \neq j}
\frac{(n_i - 1/2)\,(n_j - 1/2)}
{\left|{\bf r}_i-{\bf r}_j\right|}\,\,.
\end{equation}
Here $n_\alpha \in \{0,1\}$ denote the occupation numbers of states
localized at sites ${\bf r}_i$ within a $d$-dimensional hypercube 
of size $L^d$. Elementary charge, lattice spacing, dielectric and 
Boltzmann constants are all taken to be 1. Neutrality is achieved by 
background charges -1/2 at each site. 

For reducing finite-size effects, we impose periodic boundary conditions
for $d = 1$ and 2, using the minimum image convention \cite{Metro53}. 
For $d = 3$, the same approach would give rise to an unphysical feature:
The groundstate would be a layered arrangement of charges instead of the 
expected NaCl structure in case $L$ is a multiple of 4 \cite{Metal01}. 
Hence, for $d = 3$, we consider the sample to be surrounded by eight 
equally occupied cubes.

Our numerical investigations have been performed by means of algorithms 
based on the Metropolis procedure \cite{Metro53,Metal01,MT97}. Such 
simulations are very expensive not only for the long-range interaction, 
but also for correlation times diverging close to phase transitions and 
as temperature $T$ vanishes. Therefore it is necessary to adapt the 
dynamics to the situation that the simultaneous change of $n_i$ for 
certain clusters is necessary for overcoming passes in the energy 
landscape. To our knowledge, a procedure similar to the Swendsen-Wang 
algorithm \cite{Swen.Wang} is not available for the interaction type 
considered here. Hence we modified the dynamics ``by hand'' to take into 
account several kinds of processes: one-electron exchange with the 
surroundings, one-electron hops over ($T$ dependent) restricted 
distance, and two-electron hops simultaneously changing the occupation 
of four neighboring sites. 

At high $T$, we use the original Metropolis method \cite{Metro53}. But 
at low $T$, we take advantage of a hybrid procedure much accelerating 
the computations. It connects the direct evaluation of weighted sums 
over states within a low-energy subset of the configuration space with 
Metropolis sampling of the complementary high-energy subset \cite{MT97}.

\begin{figure}
\includegraphics[width=0.82\linewidth]{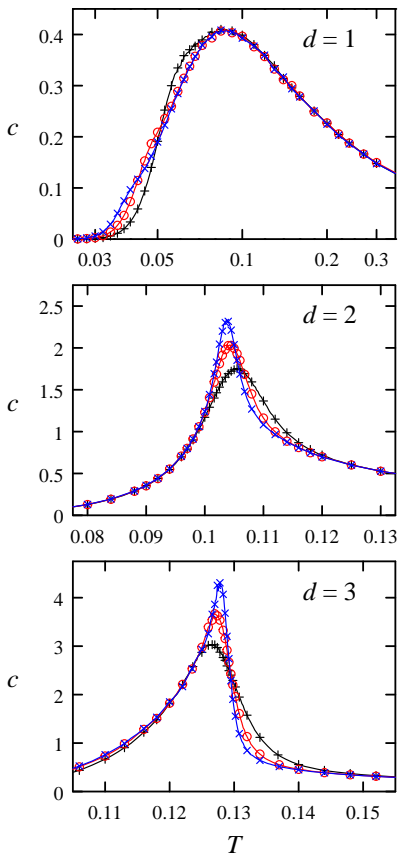}
\caption{Temperature dependences of the specific heat, $c(T)$, for 
dimensions $d = 1$ to 3 as obtained from simulations of samples of $L^d$ 
sites.
$d = 1$: $L = 100$ ($+$), 280 ($\circ$), and 700 ($\times$); 
$d = 2$: $L = 20$ ($+$), 34 ($\circ$), and 58 ($\times$); 
$d = 3$: $L = 8$ ($+$), 12 ($\circ$), and 18 ($\times$).
Only a part of the data points forming the curves is marked by symbols. 
The error bars are considerably smaller than the symbol size.}
\end{figure}

We now turn to the qualitative behavior of specific heat, order 
parameter, and susceptibility. The specific heat $c$ was obtained from 
energy fluctuations utilizing 
$c = (\langle H^2 \rangle - \langle H \rangle^2) / (T^2\ L^d) \,$.
The $T$ and $L$ dependences of $c$ are presented in Fig.\ 1 for 
dimensions $d = 1$ to 3. This graph shows that, for $d = 2$ and 3, sharp 
peaks of increasing height evolve within a small $T$ region as $L$ 
grows. Away from the peaks, within the $T$ intervals presented, $c$ is 
almost independent of $L$. But for $d = 1$, there are only broad, 
rounded peaks with $L$ independent height -- a logarithmic $T$ scale is 
used for $d = 1$, in contrast to the linear scales for $d = 2$ and 3 
which display far smaller $T$ intervals. For $d = 1$, finite size 
effects are restricted to low $T$ where the reliability bound decreases 
with $L$. 

Hence, according to the behavior of $c(T,L)$ a phase transition likely 
occurs for $d = 2$ and 3, in agreement with results of lattice gas 
simulations for $d = 3$ \cite{Letal02,DS99}. However, for $d = 1$, in 
spite of the long-range interaction, there seems to be no phase 
transition at finite $T$.

\begin{figure}
\includegraphics[width=0.82\linewidth]{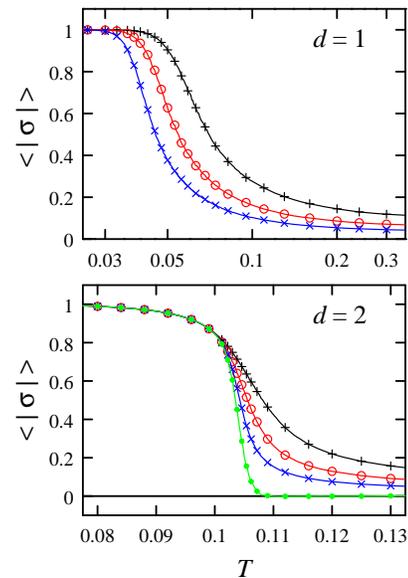}
\caption{Temperature dependence of the mean absolute value of the 
staggered occupation, $\langle |\sigma| \rangle(T)$, for $d = 1$ and 2. 
For the meaning of the symbols $+$, $\circ$, and $\times$ see caption of 
Fig.\ 1; $\bullet$ marks the extrapolation $L \rightarrow \infty$ 
explained in the text.}
\end{figure}

Analogously to an antiferromagnet, the order inherent in a charge 
arrangement $n_i$ can be characterized by means of the staggered 
occupation $\sigma_i$ relating to a NaCl structure. For example, if 
$d = 3$, it is given by
\begin{equation}
\sigma_i = (2 n_i - 1) \cdot (-1)^{x_i + y_i + z_i}
\end{equation}
where $x_i$, $y_i$, and $z_i$ denote the (integer) components of 
${\bf r}_i$. Thus we measure the mean absolute value of the staggered 
occupation $\langle |\sigma| \rangle$ as order parameter.

$T$ and $L$ dependences of $\langle |\sigma| \rangle$ are shown in 
Fig.\ 2. For $d = 1$, a rapid decrease of $\langle |\sigma| \rangle$ 
with increasing $T$ occurs already clearly below the temperature of 
maximum $c$ (same $T$ scales in Figs.\ 1 and 2). This marked decrease 
shifts to lower $T$ with increasing $L$. For $d = 2$, a qualitatively 
different behavior is found: $\langle |\sigma| \rangle$ decreases 
rapidly just in that $T$ region where the peak of $c(T)$ evolves. The 
$T$ interval of rapidly diminishing $\langle |\sigma| \rangle$ shrinks 
as $L$ rises, another indication of the phase transition. This 
interpretation is confirmed by an extrapolation $L \rightarrow \infty$: 
Assuming $\langle |\sigma| \rangle(T,L) = 
\langle |\sigma| \rangle(T,\infty) + A(T)/L^{d/2}$ with $A$ independent 
of $L$, an almost sharp transition is obtained from data for $L =34$ and 
$58$, although this extrapolation is of limited accuracy in the 
immediate vicinity of the transition. For $d = 3$, the results (not 
shown here) qualitatively resemble our findings for $d = 2$.

\begin{figure}
\includegraphics[width=0.82\linewidth]{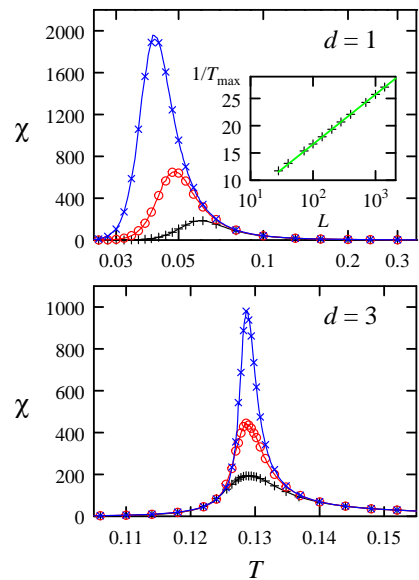}
\caption{Temperature dependence of the susceptibility, $\chi(T)$, 
related to the staggered occupation $\sigma$ by Eq.\ (3), for $d = 1$
and 3. For the meaning of the symbols see caption of Fig.\ 1. 
Additionally, for $d = 1$, the inset shows how $T_{\rm{max}}$, the 
temperature of maximum $\chi(T)$, depends on sample size $L$.}
\end{figure}

The generalized susceptibility $\chi$ related to the order parameter
$\langle |\sigma| \rangle$ is given by
\begin{equation}
\chi = L^d ( \langle \sigma^2 \rangle -
\langle |\sigma| \rangle^2) / T \ .
\end{equation}
Fig.\ 3 shows $T$ and $L$ dependences of $\chi$. It confirms the 
conclusions drawn from $\langle |\sigma| \rangle(T,L)$: On the one hand, 
for $d = 1$, a broad peak of $\chi(T)$ evolves with increasing $L$ where
$T_{\rm{max}}$, the temperature of maximum $\chi$, decreases. As the 
inset demonstrates, $T_{\rm{max}}$ can be approximated by $a/\ln(b\,L)$ 
with constants $a$ and $b$. Hence, $T_{\rm{max}}$ likely vanishes as 
$L \rightarrow \infty$ so that there seems to be no phase transition for 
$d = 1$ at finite $T$. On the other hand, for $d = 3$, as $L$ rises, a 
narrow peak grows in just that $T$ region where $c(T,L)$ has such a 
feature, a further indication of the phase transition. For $d = 2$, 
$\chi(T,L)$ behaves qualitatively similar to the results for $d = 3$.

The quantitative evaluation of our simulation data consists in a 
finite-size scaling analysis \cite{Betal00,G92}. For this aim, we first 
consider
$q_2 = -\ln(1 - \langle \sigma^2 \rangle^2 / \langle \sigma^4 \rangle)$ 
and 
$q_3 = -\tan(\pi\,(1 - 1.5 \, \langle \sigma^2 \rangle^2 / 
\langle \sigma^4 \rangle))$ 
for $d = 2$ and 3, respectively. These quantities are directly derived
from the Binder cumulant 
$1 - \langle \sigma^4 \rangle / (3 \langle \sigma^2 \rangle^2)$ and 
exhibit similar features. The $q_d(T)$ have small curvature in the 
vicinity of the transition what alleviates a precise interpolation.

\begin{figure}
\includegraphics[width=0.82\linewidth]{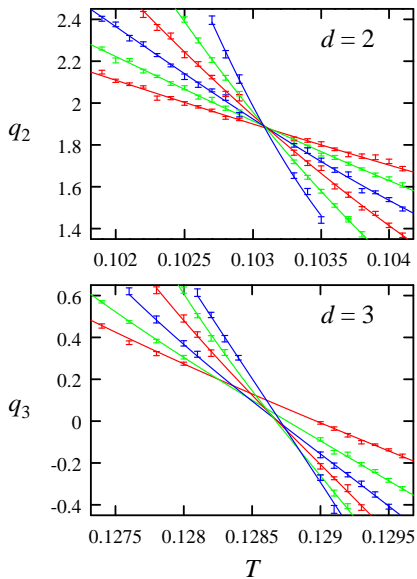}
\caption{Temperature dependence of $q_d$, related to the Binder 
cumulant, in the close vicinity of the transition for $d = 2$ and 3. For 
definitions see text. With increasing modulus of the slope, the curves 
refer to $L = 16$, 24, 34, 48, 68, and 96 for  $d = 2$, and to 
$L = 8$, 10, 12, 14, 16, and 18 for $d = 3$. For clarity, data points in 
the intersection regions are omitted.}
\end{figure}

Fig.\ 4 shows $q_d(T)$. For $d = 2$, there clearly is a common 
intersection point of the curves for different $L$ at the critical 
temperature $T_{{\rm c},2}$. But for $d = 3$, only a tendency towards 
such a behavior is seen. However, the systematic corrections to scaling 
can to a large extent be taken into account in a simple way: We  define 
a size-dependent critical temperature by 
$q_d(T_{{\rm c},d}(L),L) = q_{d,0}$ where the $q_{d,0}$ are appropriate 
constants, fixed below. Scaling of $q_d(T)$ curves for different $L$ 
with respect to $T - T_{{\rm c},d}(L)$ yields very good data collapse. 
(Referring to $t = 0$ instead of $T = T_{{\rm c},d}(\infty)$ considerably 
reduces the influence of deviations from scaling on the values for the 
critical exponents \cite{Hase}.) Thus we assume $q_{d}$ to depend only 
on $t = a_d(L) (T - T_{{\rm c},d}(L))$ in the vicinity of the 
transition.

We approximate $q_d(t)$ by polynomials of third degree, 
$q_{d,0} + t + b_d t^2 + c_d t^3$. By regression studies we first 
adjusted the $L$ independent parameters $b_d$ and $c_d$ of the ansatz, 
and then determined the $a_d(L)$ values. The latter were analyzed by 
means of power law fits including various $L$ intervals, where according 
to finite-size scaling $a_d(L) \propto L^{1 / \nu}$ was presumed. For 
consistency, the mean-square deviation of these fits must be 
understandable as resulting from random errors alone. Tab.\ I presents 
the most precise results for $\nu$, which were obtained from the fits 
safely fulfilling this requirement.

\begin{figure}
\includegraphics[width=0.82\linewidth]{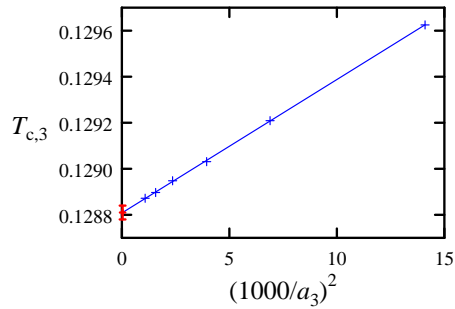}
\caption{Relation between $a_3(L)$ and $T_{{\rm c},3}(L)$ for 
$q_{3,0} = -0.174$. 
With increasing $a_3$, the points refer to $L = 8$, 10, 12, 
14, 16, and 18.  The error bar represents our extrapolation 
$L \rightarrow \infty$.
}
\end{figure}

In obtaining $T_{{\rm c},d}(L)$ from 
$q_d(T_{{\rm c},d}(L),L) = q_{d,0}$, a deviation $\delta$ of $q_{d,0}$ 
from the $L \rightarrow \infty$ limit of $q_d(T,2 L) = q_d(T,L)$ 
gives rise to a contribution 
$\propto \delta / a_d(L)$ to $T_{{\rm c},d}(L)$. We chose $q_{d,0}$ so 
that this term vanishes: $q_{2,0} = 1.8933$ and $q_{3,0} = -0.174$.  
The remaining higher order corrections in $T_{{\rm c},d}(L)$ originate
from imperfection of finite-size scaling. Comparing several 
empirical approximations, we observed that, over a 
wide $L$ range, they are almost proportional to $a_d(L)^{-2}$, see 
Fig.\ 5. Corresponding extrapolations yield the following values of 
$T_{{\rm c},d}(\infty)$: $0.10308 \pm 0.00002$ and $0.12881 \pm 0.00003$
for $d = 2$ and 3, respectively. The confidence intervals include 
the $3 \sigma$-random errors and cautious estimates for the systematic
uncertainty of the extrapolation $L \rightarrow \infty$, see Fig.\ 5.

The analysis of $c(T,L)$, $\langle |\sigma| \rangle(T,L)$, and 
$\chi(T,L)$ was performed similarly to the evaluation of $q_d(T,L)$: We 
considered $\ln c$, $\ln \langle |\sigma| \rangle$, and $\ln \chi$ as 
functions of $t$ and $L$.  For not too large $|t|$, as 
$L \rightarrow \infty$, scaling implies that each of these quantities 
is decomposable into a sum of two functions depending only on $t$ and 
$L$, respectively. However, for the $L$ regions considered here, this 
hypothesis proved to be well fulfilled only for $\ln \chi$. In the cases 
of $\ln c$ and $\ln \sigma$, there is a clear tendency towards such a 
behavior, but small deviations cannot be neglected. Thus we approximated 
$\ln c$, $\ln \langle |\sigma| \rangle$, and $\ln \chi$ by polynomials 
in $t$ of third order, taking advantage of universalities in the 
coefficients as far as possible. This regression provides precise values 
for the observables at $t = 0$. Simultaneously, we obtained the 
confidence intervals taking into account the uncertainties in the 
individual measurements of the observables and in the $T_{{\rm c},d}(L)$ 
values.

\begin{figure}
\includegraphics[width=0.82\linewidth]{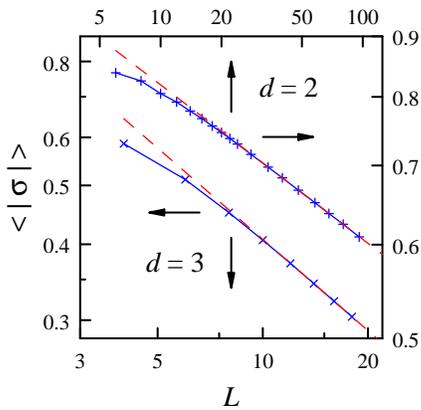}
\caption{Size dependence of the value of $\langle |\sigma| \rangle$ at 
$T_{{\rm c},d}(L)$ defined in the text. Error bars are considerably 
smaller than symbol size. The dashed lines represent the fits given in 
Tab.\ I.
}
\end{figure}

The interpolation results for $\langle |\sigma| \rangle$ and $\chi$ were 
analyzed by means of power law fits, where proportionality to 
$L^{-\beta/\nu}$ and $L^{\gamma/\nu}$, respectively, was presumed. These 
studies were performed analogously to the determination of $\nu$. Tab.\ 
I presents the values of the exponent ratios $\beta/\nu$ and 
$\gamma/\nu$. The high quality of these power law fits is illustrated by 
Fig.\ 6 presenting $\langle |\sigma| \rangle(T_{\rm c}(L),L)$.

\begin{table}
\begin{tabular}{ccccc}
Quantity & $d$ & $L$ region & Coulomb&s.-r.~Ising \\
\hline
$\alpha / \nu$&2&24 -- 96&-0.037(44)&0 (ln)\\
$\beta / \nu$&2&48 -- 96&0.131(6)&1/8\\
$\gamma / \nu$&2&48 -- 96&1.735(24)&7/4\\
$\nu$&2&34 -- 96&1.021(33)&1\\
$\alpha / \nu$&3&6 -- 18&0.09(6)&0.1761[25]\\
$\beta / \nu$&3&12 -- 18&0.499(16)&0.5181[4]\\
$\gamma / \nu$&3&14 -- 18&1.974(23)&1.9638[8]\\
$\nu$&3&10 -- 18&0.635(10)&0.6297[5]\\
\end{tabular}
\caption{Finite-size scaling results for the critical exponents of 
specific heat, staggered occupation, susceptibility, and correlation 
length, $\alpha$, $\beta$, $\gamma$, and $\nu$, respectively. To retain 
numerical precision, we mostly present exponent ratios instead of the 
exponents themselves. Values for the Ising model with short-range 
interaction {\protect \cite{G92,Hase}} are included for comparison. 
Parentheses and brackets give $3 \sigma$-random errors and total errors, 
respectively, referring to the last given digit of the value.}
\end{table}

Compared to the study of  $\langle |\sigma| \rangle$ and $\chi$, the
analysis of $c$ is more difficult: Exponent values obtained from power 
law fits converge only slowly with increasing $L$, and the mean-square 
deviations remain too large. Therefore we took into account a background 
contribution presuming
$c(T_{\rm c}(L),L) = f(L) = a + b L^{\alpha/\nu}$. However, we treated
$f(L_1)$ and $f(L_2)$ at fixed finite $L_1$ and $L_2$ as adjustable 
parameters instead of $a$ and $b$ to avoid numerical problems for almost 
logarithmic behavior of $c(L)$ (small $|\alpha|$).  Results for 
$\alpha/\nu$ obtained this way are included in Tab.\ I.

Our values in Tab.\ I have to be regarded as effective exponents. Due 
to the finiteness of $L$, tiny systematic errors are certainly present, 
presumably the more relevant the smaller the exponent value.
Unfortunately, our data set is not sufficient for a convincing estimate 
of them. However, even if only random errors are considered, our 
values comply with the Widom relation, $2 = \alpha + 2 \beta + \gamma$, 
and the hyperscaling relation, $2 - \alpha = d \, \nu$.

For comparison, Tab.\ I includes also values for the Ising model with 
nearest-neighbor interaction \cite{G92,Hase}. It is surprising that, in 
spite of the differing characters of the interactions, the critical 
exponent data obtained here are very close to these values: 
In particular, for $\gamma / \nu$ and $\nu$. the agreement is perfect 
within numerical accuracy. The slight deviations in $\alpha/\nu$ and 
$\beta/\nu$ for $d = 3$ presumably arise from a too small sample size.
Note: obtaining these values indirectly, via Widom and 
hyperscaling relation from $\gamma / \nu$ and $\nu$, yields again 
perfect agreement.

Concluding, in spite of the long-range interaction, the Coulomb system 
described by Eq.\ (1) seems to belong to the same universality class as 
the Ising model with short-range interaction. This suggests that the 
lattice Coulomb-glass model might have the same critical properties as 
the random-field short-range Ising model.

\begin{acknowledgments}
We thank H.~Eschrig, M.E.~Fisher, B.~Kramer, T.~Nattermann, M.~Richter, 
M.~Schreiber, and T.\ Vojta for helpful discussions and literature 
hints.
\end{acknowledgments}


\begin{thebibliography}{99}
\bibitem{DLR82} J.H.~Davies, P.A.~Lee, and T.M.~Rice, Phys.\
  Rev.\ Lett.\ {\bf 49}, 758 (1982).
\bibitem{V93} T.~Vojta, J.\ Phys.\ A: Math.\ Gen.\ {\bf 26}, 2883 
  (1993).
\bibitem{GY93} E.R.~Grannan, C.C.~Yu, Phys.\ Rev.\ Lett.\ {\bf 71}, 3335 
  (1993).
\bibitem{VS94} T.~Vojta, M.~Schreiber, Phys.\ Rev.\ Lett.\ {\bf 73}, 2933
  (1994).
\bibitem{Detal00} A.~D\'\i az-S\'anchez, M.~Ortu\~no, A.~P\'erez-Garrido,
  E.~Cuevas, phys.\ stat.\ sol.\ (b) {\bf 218}, 11 (2000).
\bibitem{LB02} E.~Luijten and W.J.~Bl\"ote, Phys.\ Rev.\ Lett.\ {\bf 89},
  025703-1 (2002), and refs.\ therein.
\bibitem{Letal02} E.~Luijten, M.E.~Fisher, and A.Z.~Panagiotopoulos,
  Phys.\ Rev.\ Lett.\ {\bf 88}, 184701-1 (2002).
\bibitem{Metro53} N.~Metropolis, A.W.~Rosenbluth, M.N.~Rosenbluth,
  A.H.~Teller, E.~Teller, J.\ Chem.\ Phys.\ {\bf 21}, 1087 (1953).
\bibitem{Metal01} A.~M\"obius, P.~Thomas, J.~Talamantes, and 
  C.J.~Adkins, Phil.\ Mag.\ B {\bf 81}, 1105 (2001).
\bibitem{MT97} A.~M\"obius, P.~Thomas, Phys.\ Rev.\ B {\bf 55}, 7460 
  (1997).
\bibitem{Swen.Wang} R.H.~Swendsen and J.S.~Wang, Phys.\ Rev.\ Lett.\
  {\bf 58}, 86 (1987).
\bibitem{DS99} R.~Dickman and G.~Stell, cond-mat/9906364.
\bibitem{Betal00} K.~Binder, E.~Luijten, M.~M\"uller, N.B.~Wilding, and
  H.W.J. Bl\"ote, Physica A {\bf 281}, 112 (2000), and refs.\ therein.
\bibitem{G92} N.~Goldenfeld, {\it Lectures on Phase Transitions and the
  Renormalization Group}, Frontiers in Physics, Vol.\ 85, (Wesley,
  Reading, 1992).
\bibitem{Hase} M.~Hasenbusch, Int.\ J.\ Mod.\ Phys.\ C {\bf 12}, 911 
  (2001).
\end{thebibliography}
\end{document}